\documentclass{JHEP3}

\usepackage{epsfig}

\newcommand{\be}{\begin{equation}}
\newcommand{\ee}{\end{equation}}

\title{Phases of three dimensional
large N QCD on a continuum torus}
\author{R. Narayanan
\\Department of Physics, Florida International University, Miami,
FL 33199, USA\\E-mail: \email{rajamani.narayanan@fiu.edu}}
\author{ H. Neuberger
\\ Rutgers University, Department of Physics and Astronomy,
Piscataway, NJ 08855, USA\\E-mail: \email
{neuberg@physics.rutgers.edu} }
\author{F. Reynoso
\\Department of Physics, Florida International University, Miami,
FL 33199, USA\\E-mail: \email{freyn001@fiu.edu}}

\abstract {It is established by numerical means that continuum
large N QCD defined on a three dimensional torus 
can exist in four different phases.
They are (i) confined phase; (ii) deconfined phase;
(iii) small box at zero temperature and (iv) small box
at high temperatures. 
}


\preprint{}

\begin{document}

\section{Introduction.}

Yang Mills theory on an $l^d$ continuum torus
($d>2$) exhibits a phenomenon referred to as
continuum reduction whereby the theory 
for $l > l_1 > 0$ 
is independent of $l$
\footnote{We use $l_1$ to denote the critical size
as opposed to $l_c$ since we will have a sequence
of critical sizes.}~\cite{Narayanan:2003fc,Kiskis:2003rd}.
At $l=l_1$, the theory goes from the confined phase ($0$c: $l > l_1$)
to the deconfined phase ($1$c: $l < l_1$). The order parameter is
the Polyakov loop and rotational symmetry is spontaneously
broken. More phases were conjectured to exist~\cite{Kiskis:2003rd}
in the continuum theory and these
are refereed to as $X$c phase with $X=2,\cdots, d$.

The aim of this paper is to numerically
establish the existence of the
$2$c and $3$c phase in addition to the
$0$c and $1$c phase for the continuum Yang-Mills theory
on a periodic torus. We will use the Polyakov
loop to define an order parameter 
to be labeled, $P$~\cite{Bhanot:1982sh},
and it will take values in the range [0,0.5]. 
If the $U(1)$ symmetry\footnote{The
$U(1)$ symmetry is the limit of the $Z_N$ symmetry
as $N\rightarrow\infty$.} under which the Polyakov loop
transforms non-trivially is spontaneously broken,
then $\bar P < 0.5$.
Let $\bar P_{x,y,z}$, be
the order parameters 
in the three directions. 
Then, $\bar P_x=\bar P_y=\bar P_z=0.5$
in the $0$c phase. 

There are three possibilities for the $1$c phase
and one of them is characterized by
$\bar P_y=\bar P_z=0.5$ 
and $\bar P_x < 0.5$ with rotational symmetry still
present in the $(y,z)$ plane. It is difficult to
numerically establish the order of the transition from
the $0$c to $1$c phase and we will leave it unresolved
in this paper. 

The $2$c phase also has three possibilities
and one of them is characterized by $\bar P_x=\bar P_y < 0.5$
and $\bar P_z=0.5$ with rotational symmetry present in the $(x,y)$
plane. This transition occurs when $l=l_2 < l_1$.
One can argue that the $1$c to $2$c phase transition
is first order as follows. In the $1$c phase, $\bar P_x$ was
less than $0.5$ and $\bar P_y$ was $0.5$. Since $\bar P_x=\bar P_y$ 
in the
$2$c phase, it is necessary for at least $\bar P_x$ or $\bar P_y$
to change discontinuously at the $1$c to $2$c transition.
If one operator shows a discontinuity, all operators
will generically show discontinuities and this
signals a first order transition.

Rotational symmetry is restored in the $3$c phase and
$\bar P_x=\bar P_y=\bar P_z < 0.5$. For the same reason as above,
we expect the $2$c to $3$c phase transition to be
of first order. This transition occurs when $l=l_3 < l_2$.

The $2$c phase is characterized by
two short directions and one infinitely long direction
since the theory will not depend on the length of
the direction where the U(1) is not broken. Therefore,
this phase describes large N QCD in a small box at
zero temperature (or infinite time). Confinement
cannot be addressed in the $2$c phase since we do
not have large Wilson loops. The $3$c phase 
describes large N QCD in a small box at high
temperatures. The $2$c to $3$c transition is
like the transition seen in perturbation theory
on $S^2\times S^1$\cite{Papadodimas:2006jd}
where $S^2$ replaces the two torus along which
the U(1) symmetry is broken in the $2$c phase.

We extend our discussion to include $l_x \times l_y \times
l_z$ torus with $l_x < l_y < l_z$. 
The transition from $0$c to $1$c will occur at $l_x=l_1$
and this is independent of $l_y$ and $l_z$. 
The transition from $1$c to $2$c will occur at $l_y=l_2(l_x)$
with $0 \le l_x \le l_2$. Furthermore, $l_2(l_2)=l_2$
and our numerical results will show that $l_2(0) > 0$. 
Finally, there is no dependence on $l_z$.
Continuing along the same lines, we can say that the
$2$c to $3$c transition occurs at $l_z=l_3(l_x,l_y)$
with $l_3=l_3(l_3,l_3)$. It is possible that one
can obtain this critical size for small
$l_x$ and $l_y$ by perturbation theory
but it is necessary to consider the zero momentum modes
of the gauge fields on all three directions. We do
not address this problem in the paper.

The results in this paper complement the results in the
closely related paper by Bursa and Teper~\cite{Bursa:2005tk}.
We mainly focus on the continuum limit of the various phases
using Polyakov loops. The paper
is organized as follows. We define the relevant technical
details in section~\ref{latdef}. Our
numerical results showing the existence of the
various phases are presented in section~\ref{polydet}.

\section{Technical details \label{latdef}}
\subsection{Lattice gauge action}
We used the single plaquette Wilson action given by
\begin{eqnarray}
S=\frac{\beta}{4N}\sum_{x,i\ne j} Tr[ U_{ij}(n)
+U_{ij}^\dagger (x) ] \\
U_{ij}(n)=U_i (n) U_j (n+\hat i) U_i^\dagger (n+\hat j) 
U_j^\dagger (n)
\end{eqnarray}
$n$ is a three component integer
vector labeling the site, $i$ labels a direction and
$\hat i$ denotes
a unit vector in the $i$ direction. The link matrices $U_\mu(n)$ 
are in $SU(N)$.
We define 
\be
b=\frac{\beta}{2N^2}
\ee
and take 
the large $N$ limit with $b$ held fixed.

All computations were done on a 
$L_x \times L_y \times L_z$ periodic lattice with $L_x \le L_y = L_z$.
One gauge field update of the whole lattice~\cite{Kiskis:2003rd} 
is one Cabibo-Marinari heat-bath
update of the whole lattice 
followed by one SU(N) over-relaxation update of the whole lattice.
The code was run on two clusters, one with 48 nodes and another
with 31 nodes. The nodes in the cluster were simply used to
generate more statistics using a parallel random number generator
and generating independent configurations with the same set of
parameters on different nodes.

\subsection{Determination of the critical sizes}
Given the lattice coupling $b$ and lattice sizes $L_x$ and $L_y$,
the dimensionless physical sizes are defined as 
\be
l_{x,y}=\lim_{b\rightarrow\infty} L_{x,y}/b_{\rm tad}.
\ee
The tadpole improved coupling~\cite{Lepage:1996jw},
$b_{\rm tad}$ is defined as
\be
b_{\rm tad} = b e(b) = b \langle
\frac{1}{12NL_x L_y L_z}\sum_{n,i\ne j} Tr[ U_{ij}(n)
+U_{ij}^\dagger (n) ] \rangle \label{plaq}
\ee

\subsection{Lattice bulk transition}
Since the computations in this paper use the Wilson gauge action
on the lattice, it is necessary to address the
unphysical transition which is the extension
of the Gross-Witten transition~\cite{Gross:1980he} in QCD${}_2$.
The order parameter for this transition is the plaquette
operator. The third order transition analytically computed
in QCD${}_2$ remains to be true based on a numerical
investigation in QCD${}_3$~\cite{Bursa:2005tk}
and the critical point is $b=0.43$.
This lattice transition does not survive the continuum
limit and we will work with $b>0.43$ throughout this paper
in order to describe continuum physics.

\subsection{An order parameter}
An order parameter suitable for studying the phase transitions
we are interested in is~\cite{Bhanot:1982sh}
\begin{eqnarray}
\bar P_{x,y,z} &=& \left < P_{x,y,z} \right > \cr
P_{x,y,z} &=& \frac{1}{2 L_x L_y L_z} \sum_n  1 - \left | \frac{1}{N} 
Tr {\cal P}_{x,y,z}(n)
\right |^2 \cr
{\cal P}_{x,y,z}(n) &=& \prod_{m=1}^{L_{x,y,z}} U_i(n+m\hat i).
\end{eqnarray}
The quantity $P_{x,y,z}$ takes values in the range $[0,0.5]$ on
any gauge field background and we choose the 
$x$, $y$ and $z$ 
directions on each configuration
such that $P_x < P_y < P_z$.

Although this observable needs to be renormalized, we found it
sufficient to work with the unrenormalized operator and
we also did not have to smear the link variables.
The eigenvalues,$e^{i\theta_k}$; $k=1,\cdots, N$,
 of the Polyakov loop operator,
${\cal P}_{x,y,z}(n)$, are gauge invariant.
$P_{x,y,z}=0.5$ implies a uniform distribution of the eigenvalues
of ${\cal P}_{x,y,z}(n)$.
A departure from $P_{x,y,z}=0.5$ implies the presence of
a peak in the distribution of the eigenvalues of ${\cal P}_{x,y,z}(n)$
and a breaking of $Z_N$ symmetry associated with this operator.
There is no gap in the distribution of the eigenvalues
of ${\cal P}_{x,y,z}$ when $Z_N$ is broken.

\section{Transitions in Polyakov loops \label{polydet}}

All computations were done using $N=47$. Having picked a lattice
size $L_x \times L_y \times L_y$, each run listed in Table~\ref{tab1}
was a closed loop in the lattice coupling $b$. 
The third column in Table~\ref{tab1}
shows the range of $b$ and the step size in $b$. The fourth
column, $N_t$, shows the number of thermalization 
sweeps at the two end points. Only one measurement was done per node
at each $b$ and
the fifth column, $N_b$, shows the
number of sweeps done at each intermediate $b$ before the measurement.
For example,
the run on $3^3$ lattice, 
started at a $b=0.5$ and went up to a $b=2.5$.
A total of $2000$ sweeps were performed at $b=0.5$ and $b=2.5$ and
a total of $1000$ sweeps were performed for all $b$ in between $0.5$
and $2.5$. The step size in $b$ was $0.05$ and this code was run on
the 31 node cluster with one measurement at each $b$ per node. All
values of $b$ between $0.5$ and $2.5$ had two sets of measurements;
one on the way up in $b$ and one the way down in $b$. 

\TABLE{
\begin{tabular}{ccccccccc}
$L_x$ & $L_y$ & $b$ & $N_t$ & $N_b$ & $N_{\rm cfg}$ 
& $\frac{L_x}{b_{\rm tad}}(0{\rm c}-1{\rm c})$ 
& $\frac{L_y}{b_{\rm tad}}(1{\rm c}-2{\rm c})$ 
& $\frac{L_z}{b_{\rm tad}}(2{\rm c}-3{\rm c})$ \cr
\hline
3 & 3 & [0.5,2.5;0.05] & 2000 & 1000 &  31 & 6.14(33) & 4.02(14) & 2.27(17)\\
4 & 4 & [0.5,3.5;0.05] & 3000 & 600 &  48 & 5.76(21) & 3.83(28) & 2.14(17)\\
5 & 5 & [0.5,2.5;0.05] & 2000 & 400 &  31 & 5.60(47) & 3.73(35) & 2.17(23)\\
6 & 6 & [0.5,4.5;0.10] & 2000 & 400 &  31 & 5.37(48) & 3.82(70) & 1.99(38)\\
5 & 6 & [1.5,3.5;0.10] & 3000 & 600 &  48 & & 2.70(25) & \\
4 & 5 & [0.5,3.5;0.05] & 3000 & 600 &  48 & 6.00(46) & 2.47(23) & \\
3 & 4 & [0.5,2.5;0.05] & 3000 & 600 &  48 & 6.56(76) & 2.07(20) & \\
4 & 6 & [2.0,5.0;0.10] & 3000 & 600 &  48 &  & 1.78(17) & \\
3 & 5 & [2.0,4.0;0.10] & 3000 & 600 &  48 &  & 1.53(15) & \\
3 & 6 & [2.0,8.0;0.20] & 2000 & 400 &  31 &  & 1.14(11) & \\
\hline
\end{tabular}
\caption{\label{tab1} The parameters of all the runs used
to study the transitions in Polyakov loops along with the
results for the critical sizes.}}

\subsection{Details of the data analysis}

The plaquette
as defined in (\ref{plaq}) was measured on all configurations and
this was used to obtain the tadpole improved coupling, $b_{\rm tad}$.
Figure~\ref{444} shows the results for all three Polyakov loop
observables as a function of $\frac{4}{b_{\rm tad}}$ for the data
obtained on the $4^3$ lattice (second row in Table~\ref{tab1}).
The hysteresis is clear in both
$\bar P_x$ and $\bar P_y$ for the $1$c-$2$c transition and 
it is seen in all the $\bar P_{x,y,z}$ 
for the $2$c-$3$c transition. 
The critical size for the various transitions along with the
error is obtained by locating the two points 
(one for upward direction and the second for the
downward direction) where the error
is largest in the observable that is broken. The
vertical lines in Figure~\ref{444}
shows the critical sizes along with the errors
and these results are shown in the 
last three columns of Table~\ref{tab1}. 

Within the $2$c phase one sees a difference in $\bar P_{x}$
and $\bar P_y$. But this is just a consequence of our
choice of observable. Note that we have picked $P_x < P_y$
on every configuration. If we assume two
independent Gaussian random variables, $\alpha$ and $\beta$,
that have the same mean and variance, then one can
show that the variables $P_x$ and $P_y$ defined as the
minimum and maximum of $\alpha$ and $\beta$ will
be distributed such that 
\be
\frac{ \bar P_y - \bar P_x} 
{\sqrt{ \left< P^2_{x,y} \right> - P^2_{x,y} }} = 
\frac{2}{\sqrt{\pi -1}}.
\ee
Our data within the $2$c phase is consistent with the above
equation.

We did not choose a range in $b$ such that all transitions
are seen on all $L_x$, $L_y$ pairs since some of them were
used only to investigate the $1$c-$2$c transition. But,
we always picked a range such that the end points are
in one of the four phases.
 
\FIGURE[htp]{
\epsfig{file=poly_444_600.eps, height=3.5in }
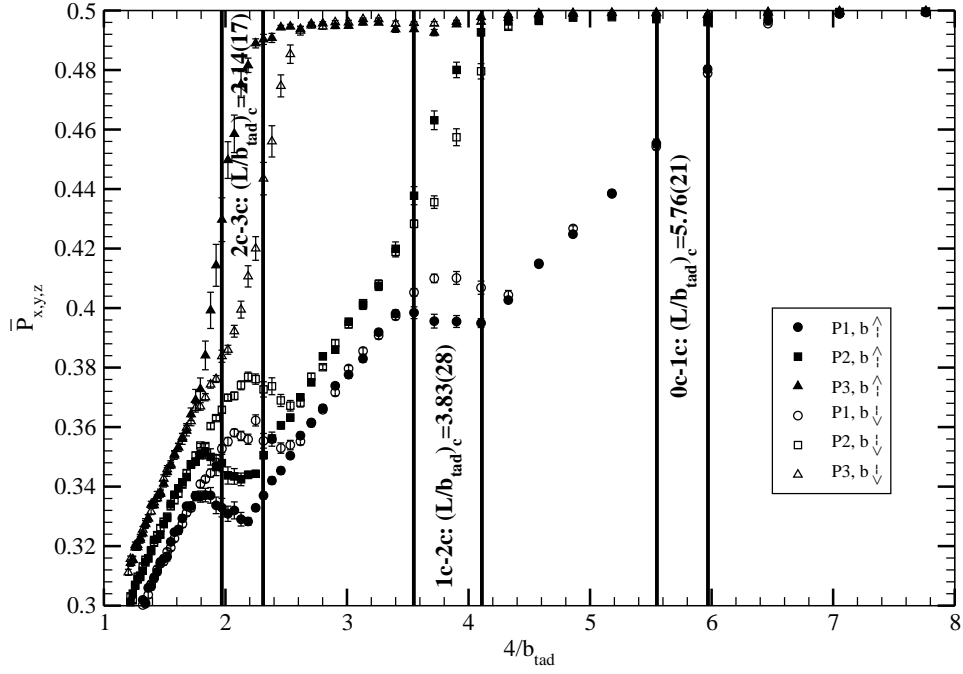
\caption{Plot of $\bar P_{x,y,z}$ for the data
in the second row of Table~\ref{tab1}
showing all three transitions.}
\label{444}}

\subsection{$0$c-$1$c transition}

Let us focus on the seventh column in Table~\ref{tab1}
to study the confinement-deconfinement transition. The results on
$3^3$, $4^3$, $5^3$ and $6^3$ show that the $0$c-$1$c transition
is physical since the critical size, $\frac{L_x}{b_{\rm tad}}$,
is the same on all four lattices within errors. The results
here are consistent with the older results presented 
in~\cite{Narayanan:2003fc}. We also studied the $0$c-$1$c transition
on $4\times 5^2$ and $3\times 4^2$ and found that the critical
size is independent of $L_y$ as expected. Figure~\ref{P1} shows
that the six results for the $0$c-$1$c transition do scale properly
and we estimate the continuum critical size to be
$l_1=5.90(47)$. 
If we take the central value for the dimensionless string tension 
from~\cite{Bringoltz:2006zg}, namely, $\sqrt{\sigma}=0.1975$,
then we get 
\be
\frac{1}{l_1\sqrt{\sigma}}=0.86(7)
\ee
and this is consistent~\cite{Liddle:2005qb}
 with saying that $\frac{1}{l_1}$
is the deconfinement temperature.

\FIGURE[htp]{
\epsfig{file=P1.eps, height=3.5in }
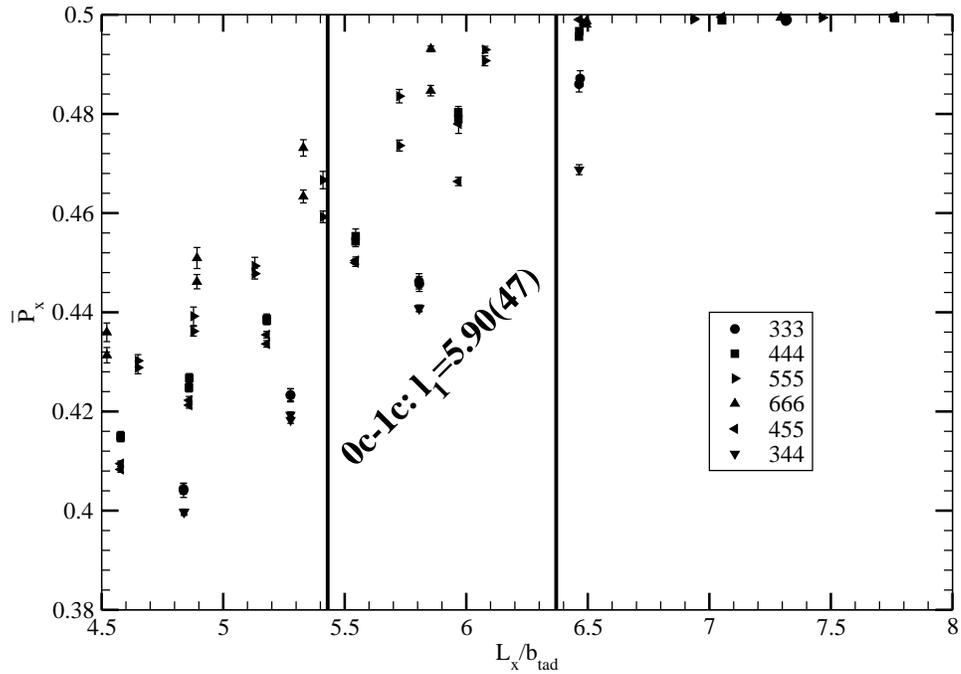
\caption{Plot of $\bar P_x$ showing the $0$c-$1$c transition.}
\label{P1}}

\subsection{$1$c-$2$c transition on $L^3$ lattices}\label{1to2}

The physical size associated with the $1$c-$2$c transition, $l_2$.
is expected to depend on $l_x$, the
temperature in the deconfined phase. 
We first estimate the critical
size on lattices with $L_x=L_y$. We use the data on
$3^3$, $4^3$, $5^3$ and $6^3$.
The four results show continuum scaling as can be seen
from Figure~\ref{P2a} and we conclude that
the $1$c-$2$c transition exists in the continuum limit. 
We estimate $l_2(l_2)=3.85(43)$.
As mentioned before, $1$c phase is the deconfined phase.
The system is a small finite box at zero temperature
in the $2$c phase. This transition occurs on a
$l^3$ torus when the temperature is $1.53(21)$ times
the deconfinement temperature.

\FIGURE[htp]{
\epsfig{file=P2a.eps, height=3.5in}
\caption{Plot of $\bar P_y$ showing the $1$c-$2$c transition
on lattices with $L_x=L_y$.}
\label{P2a}}

\subsection{$2$c-$3$c transition on $L^3$ lattices}

We also investigated the $2$c-$3$c transition on
$3^3$, $4^3$, $5^3$ and $6^3$.
Here again,
the four results show continuum scaling as can be seen
from Figure~\ref{P3} and we conclude that
the $2$c-$3$c transition also exists in the continuum limit. 
The transition size will depend on $l_x$ and $l_y$ when
both of them are smaller than $l_z$. But, we only estimate
$l_3(l_3,l_3)=2.14(26)$ here.

Large N QCD on a very small torus, $l^3$,
for $l< l_3$ feels
the size of the box and the temperature is high.
Large N QCD is in a small box of size $l$ 
at zero temperature if $l > l_3$ and
it
undergoes a phase transition into the deconfined phase
when the box
size is $1.80(30)$ times $l_3$.

\FIGURE[htp]{
\epsfig{file=P3.eps, height=3.5in}
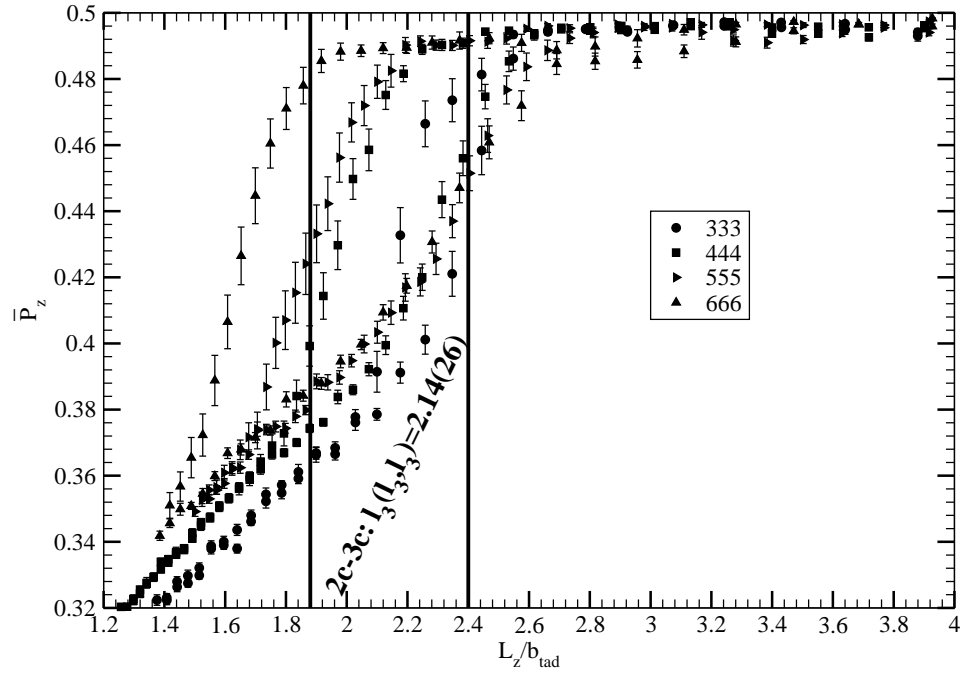
\caption{Plot of $\bar P_y$ showing the $2$c-$3$c transition
on lattices with $L_x=L_y$.}
\label{P3}}

\subsection{Phase diagram for $l_x \le l_y \le l_z$}

The single scale in the $1$c phase is $l_x$ which
can also be thought of as inverse temperature in the
deconfined phase. The $2$c phase has two scales,
namely the size of the two dimensional box $l_x$ and $l_y$
with $l_x \le l_y$.
If $l_y > l_2(l_x)$, then the theory does not depend on
$l_y$ and we are in the deconfined phase. We considered
the special case of $l_x=l_y$ in section~\ref{1to2}.
We extended this to the case when $l_x < l_y$.
For this purpose,
we considered the lattices listed in last six rows of Table~\ref{tab1}.
The phase transition in $\bar P_y$ is shown in Figure~\ref{P2b}. 
There is an
obvious dependence of the critical size $l_2$ on $l_x$.

Figure~\ref{xvsy} summarizes the various phases by focusing
on the $(l_x,l_y)$ plane at $l_z=l_y$.
The dependence of $l_2(l_x)$ is shown using the shaded
square points in Figure~\ref{xvsy}. The dashed line is
a quadratic fit to the seven points and we note that
$l_2(0) > 0$.
In order to get an overall picture, we have also shown
the $0$c-$1$c transition in Figure~\ref{xvsy}. The
dotted line indicates that the $0$c-$1$c transition
does not depend on $l_y$ for $l_y > l_x$. Figure~\ref{xvsy}
also shows the $0$c, $1$c and $2$c phases for $l_x \le l_y \le l_z$.
For completeness, we have also shown the $2$c-$3$c transition
as seen on this specific $(l_x,l_y)$ plane restricted to $l_x=l_y$.
Like the $1$c-$2$c transition, the $2$c-$3$c transition will also
show a dependence on $l_x$ for $l_x< l_y$
and the $2c$ phase will not reach the $l_x=l_y$ line
for $l_x=l_y < l_2$. Furthermore, the
$2$c-$3$c transition curve will change as one changes the
$l_z$ that defines the $(l_x,l_y)$ plane. We have not
investigated these details pertaining to the $2$c-$3$c
transition in this paper. But, we should remark that the
rest of the phase diagram does not depend on $l_z$ for
$l_z > l_y > l_x$.

\FIGURE[htp]{
\epsfig{file=P2b.eps, height=3.5in}
\caption{Plot of $\bar P_y$ showing the $1$c-$2$c transition
on lattices with $L_x \le L_y$.}
\label{P2b}}

\FIGURE[htp]{
\epsfig{file=xvsy.eps, height=3.5in}
\caption{Phase diagram for $l_x \le l_y \le l_z$}
\label{xvsy}}

\section{Conclusions}

Large $N$ QCD in three dimensions on a $l^3$
continuum torus exists in four different phases. The theory
is in the confined phase ($0$c) for $l \ge 5.90(47)= l_1$
and physics does not depend on the box size. This critical
size is the inverse of the deconfinement temperature, 
$T_c=\frac{1}{l_1}$, 
and the
theory is in the deconfined phase ($1$c) for 
$1 < \frac{T}{T_c} < 1.53(21)$.

The system is in a finite box and feels the
effect of temperature ($3$c phase) when $l  < 2.14(26)= l_3$.
The temperature has no effect if $1 < \frac{l}{l_3} < 1.80(30)$
($2$c phase).
The system goes into the deconfined phase if
$\frac{l}{l_3} > 1.80(30)$. 

All phase transitions are most likely first order in nature.
We have provided arguments for this scenario when going
from  $1$c-$2$c and $2$c-$3$c. 
Previous results~\cite{Liddle:2005qb,firstorder} 
indicate that the deconfinement phase transition is also first order.

The $1$c to $2$c transition on a $l_x \times l_y \times l_z$ torus
with $l_x \le l_y \le l_z$ depends on $l_x$. The critical line is
given by $l_2(l_x)=0.56 + 1.08 l_x - 0.059 l_x^2$ and this is valid
for $0 \le l_x \le 3.85(23)$ and it is independent of $l_z$. This
transition has been analyzed for one point on the $(l_x,l_y)$ plane
in~\cite{Bursa:2005tk} and we are in agreement with the result in that
paper.

\acknowledgments

R. N. and F.R. acknowledge partial support by the NSF under grant number
PHY-055375 
R.N. also acknowledges partial support from Jefferson Lab. The Thomas Jefferson National
Accelerator Facility (Jefferson Lab) is operated by the
Southeastern Universities Research Association (SURA) under DOE
contract DE-AC05-84ER40150. H. N. acknowledges partial
support by the DOE under grant number DE-FG02-01ER41165 at Rutgers,
an Alexander von Humboldt award and the hospitality of the Physics
department at Humboldt University, Berlin.

\end{document}